\documentclass[12pt]{article}

\usepackage[dvips]{graphicx,color}

\begin{document}

\begin{center}

{\Large {\bf Higgs mechanism from fluxes and \\

\vspace{0,3cm}

two mass hierarchies in the "fat" throat\\

\vspace{0,3cm}

solution of Type IIA supegravity}}

\vspace{2cm}

{Boris L. Altshuler}\footnote[1]{E-mail adresses: altshuler@mtu-net.ru \& altshul@lpi.ru}

\vspace{0,5cm}

{\it Theoretical Physics Department, P.N. Lebedev Physical
Institute, \\  53 Leninsky Prospect, Moscow, 119991, Russia}

\vspace{2cm}

{\bf Abstract}

\end{center}
      
Spectra of Kaluza-Klein gauge fields are calculated at the background of the magnetic fluxbrane throat-like solution of the Type IIA supergravity equations. Magnetic flux plays the role of Higgs scalar generating the electroweak scale mass of non-abelian KK gauge field. The same "flux-Higgs" mechanism gives the scale of the "second mass hierarchy", $10^{-3} eV$, in the spectrum of KK excitations at the background of the throat deformed in a Reissner-Nordstrom way. 
Gauge coupling constants in 4 dimensions are 
calculated; they prove to be 
of the physically sensible values for the EW scale massive modes of the 
gauge fields and are extremely small for the long range gauge fields.
5D effective Action with the flux-generated mass terms of the KK gauge 
fields associated with isometries of compactified subspaces is put down, 
correspondence of the "classical" KK approach and dual holography 
approach to the low-dimensional gauge theories is discussed, and idea 
of {\it{bootstrap holography}} is set forth.

\vspace{0,5cm}

\newpage
\section{Introduction. Overview of approach and results}

\qquad In \cite{Altsh7}, \cite{Altsh6} radion effective potential compatible 
with inflation scenario was received in the model of magnetic fluxbrane 
throat-like solution of the Type IIA supegravity equations. 
Radion ("inflaton") field is defined in \cite{Altsh7}, \cite{Altsh6} as 
depending on 4 space-time coordinates position of the 
positive-tension isotropic brane which fix the UV end of the throat. 
Radion potential possesses wide relatively flat region for inflation, 
steep slope for reheating and stable extremum determined by the Israel 
jump conditions at the mouth of the throat 
where Universe resides in modern epoch. In the present paper we 
take this stable solution as a background and explore, partly, the 
physical content (spectra, gauge coupling constants) of its Kaluza-Klein 
excitations. Thus we cut out the Ricci-flat Calabi-Yau bulk 
and consider only inner region of the background throat.

To get the discrete spectrum and the mass gap the IR end of the throat 
must be fixed. To make it dynamically the Euclidian version of the 
Reissner-Nordstrom (RN) deformation of the 
elementary magnetic fluxbrane solution is used. It was shown in \cite{Altsh7} 
that this results in shift of the extremal 
value of radion potential from zero to small positive quantity 
$m_{EW}^{8}/M_{Pl}^{4}$ ($m_{EW} \cong 1 TeV$ is the electroweak scale, 
$M_{Pl}=2.4 \times 10^{18} GeV$ is the reduced Planck mass), which is of 
order of vacuum energy responsible for the observed acceleration of the 
Universe and which for unknown reason is equal to $m_{\nu}^{4}$, 
$m_{\nu}=10^{-3}eV$ is mass of light neutrino. $m_{EW}/M_{Pl}= 10^{-15}$ is 
the first mass hierarchy, whereas $m_{\nu}/M_{Pl}=m_{EW}^{2}/M_{Pl}^{2}=10^{-30}$ 
is so called "second mass hierarchy". In the present paper this small mass 
scale appears in other context.

Schwarzschild type or RN type deformations of the throat-like 
elementary solutions always demand introduction of the additional cyclic 
dimension which plays the role of Euclidian "time". The study of spectra of 
KK excitations of gravity and matter fields on the supergravity background 
of strongly warped space-time with additional extra cyclic dimension (so 
called "AdS-soliton" or "D4-soliton" solutions) possesses a good history, 
\cite{Witten} - \cite{Myers8} and references therein. In many of these papers 
spectrum of glueballs and other physical quantities of the dual gauge CFT 
theory on the boundary of higher dimensional warped space-time were calculated 
in frames of the gauge/gravity duality based upon the idea of the AdS/CFT 
correspondense and holography \cite{Maldacena}-\cite{Aharony}. Also 
excitations on the background of the Klebanov-Strassler (KS) 
warped deformed conifold 
\cite{Klebanov} were studied in the context of attempts to receive some 
features of QCD as a gauge theory dual to supergravity (see e.g. 
\cite{Krasniz}-\cite{Klebanov07}). However much work is to be done in this 
direction to make sure that holographic dual gauge theories fit for the 
description of real world \cite{Myers8}.

Here we don't touch this most interesting field of research and consider the 
"classical" Kaluza-Klein approach \cite{KK} to gauge theories when gauge 
potentials are the non-diagonal components (vectors in lower dimensions) of 
the higher-dimensional metric tensor. There were many hopes to receive the 
realistic theory of electromagnetic and other observed fields by 
compactification of the higher-dimensional gravity \& supergravity 
\cite{AnnPhys}, \cite{PhysRep}, \cite{SUGRA}. However 
this dreamed for goal is still "behind the horizon". Results of this paper 
are hopefully the step in this direction on the methodological level of the 
toy model.

The possible correspondence of two promising approaches (dual one and KK one) 
to the low-dimension gauge theories is a special challenging issue which will be shortly 
discussed in the last Section of the paper.

We explore the KK gauge theory resulting from compactification from 10 
to 5 (and then to 4) dimensions when deformed throat-like solution 
of the Type IIA supergravity is taken as a background. 
Schredinger-type equations 
determine the spectra of excitations in a way which may be applied to 
any model of a throat, including e.g. KS throat of the Type IIB supergravity.
Received spectra qualitatively coincide with 
mass spectra of glueballs calculated in \cite{Witten} - \cite{Myers8}, 
\cite{Krasniz}-\cite{Klebanov07}. However 
approach of this paper looks quite simple and general. 

The dynamical construction of the 
smooth IR tip of the throat is a great advantage of the model of deformed 
KS throat \cite{Klebanov} and of the models with extra cyclic dimension. In 
most of the models with additional cyclic dimension the closure of the bulk 
regularly at codimension two submanifold known as "bolt" \cite{Hawking} is 
achieved with use of the Euclidean version of the Schwarzschild type 
deformation of the elementary throat-like solution. But the 
RN type deformation also may be used for this purpose, 
see e.g. \cite{Gibbons}, \cite{Louko}, \cite{Burgess3} where exact 
solutions in $p+2$ dimensions and in 6 dimensions were considered.
Contrary to these models the D10 RN-deformed throat is an approximate 
solution of the dymanical supergravity equations; in \cite{Altsh7} it was 
shown however that deviations from the exact solution prove to 
be not too essential if the throat is sufficiently long as compared to the Planck 
scale, which in turn is demanded by the large value of the mass hierarchy 
calculated in the model. 

Like in the Randall and Sundrum (RS) model \cite{RS} we ajust location of the 
IR end of the warped space-time to get the needed value of the first mass 
hierarchy $m_{EW}/M_{Pl}=10^{-15}$. At the same time the model under 
considerartion does not suffer from known shortcomings of the RS model 
which is essentially built upon the idea of trapping of the SM matter 
upon the defect solutions \cite{Rubakov}, \cite{Akama} and where because 
of it the asymmetric treatment of the bulk and brane physics, of gravity 
and matter is at hand \cite{Benson}. 
Eventually it became clear that role of "trapping" may 
play the gravitational accretion of massive matter to the IR end of the 
strongly warped space-time. Thus there is no need in special tools of trapping, 
no need to make assumptions about primary masses of matter fields trapped on 
a brane. Wave functions of massive excitations of bulk fields are concentrated 
in vicinity of the IR end. Because of it and also because zero mode of gravity 
is diluted uniformly in the extra space the RS tool of getting mass hierarchy 
is realized automatically for the massive KK excitations of the bulk fields.

Thus in the paper we'll consider only bulk fields without artificially 
added mass terms in the higher-dimensional matter Lagrangian. Taking 
mass gap in KK spectrum of excitations to be equal to $m_{EW}$
determines parameters of the throat: width of the throat 
(given by the value of period of cyclic dimension at the UV end of the throat 
which actually fixes 
the value of the mass gap, cf. \cite{Wolfe02}) must be taken of the 
electroweak scale $\cong 10^{-16} sm$; then proper length of the throat 
proves to be 6 orders less. That is why we call this construction the "fat" throat.

In the paper the mechanism of exclusion of zero mass modes from spectrum of 
KK gauge excitations is considered which does not suppose introduction of 
the fundamental or composite Higgs scalar. Alternatives to Higgs mechanism 
were considered earlier, see e.g. \cite{Shaposh} and references therein. 
Here we propose another option which actually is not alternative 
to Higgs mechanism, but it is shown that instead of the Higgs scalar the 
magnetic flux, i.e. $n$-form which "lives" on the compact 
submanifold "responsible" for the KK 
gauge field at hand may be used. Then non-zero background components of the 
$n$-form field (which evidently are not invariants of the gauge transformations 
from the isometry group of this manifold) play the role of vacuum condensate of 
traditional charged Higgs scalar. In this way the 
EW mass of the non-abelian KK gauge field associated with isometry group of 
the D4 compact subspace of the model is naturally received. The unexpected 
"side-effect" of the proposed "flux-Higgs" mechanism is appearance of the 
"second mass hierarchy" value $m_{EW}^{2}/M_{Pl}=10^{-3}eV$ in the spectrum 
of KK gauge field associated with extra cyclic dimension. 

The difficulty of application of traditional KK approach to the throat-like 
models of supergravity theories is in the values of the effective, in 4 
dimensions, gauge coupling 
constants of KK gauge fields. Calculations show that for zero modes, which are 
uniformly diluted in extra space, gauge coupling constants are extremely small 
because of the big size of the throat as compared to the Planck scale. Thus 
these modes can not be observed (cf. mechanism of exclusion of the gauge field 
zero modes from the physical sector proposed in \cite{Shaposh}). However, 
contrary to zero modes, the calculated values of the gauge coupling 
constants of the massive modes of KK gauge fields prove to be physically 
acceptable in the throat-like models of the Type IIA supergravity. But it is 
not the case for the similar models of the Type IIB supergravity (see Sec. 6).

The paper is organized as follows. After description of the background 
solution (Sec. 2) the simple illustrative example of calculation of spectrum 
of excitations of the bulk zero mass scalar field is considered in Sec. 3. In 
Sec. 4 dynamics of the KK gauge fields received by compactification from 10 
to 5 dimensions is considered, the work of the "flux-Higgs" mechanism is 
demostrated and spectra of eigenmodes of gauge fields is calculated. 
In Sec.5 appearance of the "second" mass hierarchy is shown, and in Sec. 6 
gauge coupling constants for different gauge fields' eigenmodes are calculated. 
In Sec. 7 results are discussed.

\section{Description of the "fat" throat}

We use the truncated bulk Action of the Type IIA supergravity in the Einstein frame:

\begin{eqnarray}
\label{1}
&&S^{(10)}=M^{8}\int\left[R^{(10)}-\frac{1}{2}(\nabla\varphi)^2-\frac{1}{48}e^{\varphi /2}F_{(4)}^2  \right.  \nonumber   
\\
&&\left. -\frac{1}{4M^{2}}e^{3\varphi/2}F_{(2)}^{2} \right]\sqrt{-g^{(10)}}\,d^{10}x+\rm{GH},
\end{eqnarray}
where $M$, $g_{AB}$, $R^{(10)}$ are Planck mass, metric and curvature in 10 dimensions ($M\equiv M_{(10)}=l_{s}^{-1}g_{s}^{-1/4}$, $l_{s}$, $g_{s}$ are string length and string coupling constant, $A,B = 0,1...9$); $\rm{GH}$ - Gibbons-Hawking term; $F_{(4)}=F_{ABCD}$ is the 4-form field strength; $F_{(2)}=F_{AB}$ - 2-form field conventionally normalized, $F_{(4)}^{2}=F_{ABCD}F^{ABCD}$, $F_{(2)}^{2}=F_{AB}F^{AB}$, $\varphi$ - dilaton. Let us take as a background the extremal magnetic fluxbrane throat-like solution of theory (\ref{1}) deformed in a "Reissner-Nordstrom" way \cite{Altsh7}:

\begin{eqnarray}
\label{2}
&&ds_{(10)}^{2}=b^{2}\tilde{g}_{\mu\nu}dx^{\mu}dx^{\nu}+ c^{2}d\theta^{2}+N^{2}dr^{2}+a^{2}g_{ij}dy^{i}dy^{j},
\\  \nonumber
&& b=H^{-3/16}, \quad c=\frac{T}{2\pi} U^{1/2}H^{-3/16}, \quad N= U^{-1/2}H^{5/16}, \quad a=r\,H^{5/16}, 
\\ 
&& e^{\varphi}= g_{s} H^{-1/4}, \qquad H=1+\left(\frac{L}{r}\right)^{3}, \qquad U=1-\left(\frac{l}{r}\right)^{6},  \nonumber
\\ \nonumber
&& F_{(4)}=3L^{3}g_{s}^{-1/4}\epsilon_{4},  \qquad F_{(2)\theta r}=3\sqrt{2}\,M\, l^{3}g_{s}^{3/4}\frac{cN}{b^{4}a^{4}}e^{-3\varphi/2}  \nonumber
\end{eqnarray}
where $\tilde{g}_{\mu\nu}=\eta_{\mu\nu}=\it{diag}(-+++)$;  $g_{ij}$ is metric of compact 4-dimensional sub-space $K$, $\epsilon_{4}$ is its 4-form; $\mu,\nu = 0,1,2,3$; $i,j = 1,2,3,4$; $r$ is isotropic coordinate, $\theta$ ($0<\theta<2\pi$) is additional cyclic coordinate with period $T$ which plays the role of Euclidian "time" in the Reissner-Nordtstrom type deformation of the elementary throat-like solution permitting to fix smoothly the IR end of the throat at the point $U=0$. The characteristic length $L$ of the throat in the string-brane vocabulary is $L^{3}=g_{s}N_{c}l_{s}^{3}$ where $N_{c}$ is number of branes in a stack of branes forming the throat (\ref{2}) or equivalently the number of colors of dual gauge theory. (Cf. \cite{Myers8} where solution (\ref{2}) is written down inside the throat in string frame).

Contrary to the often used Schwarzschild type deformation of the elementary fluxbrane throat, when in (\ref{2}) $U=1-l^{3}/r^{3}$ which is the exact solution of the dynamical equations of Type IIA supergravity, the Reissner-Nordstrom type deformation given in (\ref{2}) is an approximate solution of these equations \cite{Altsh7} (see comment in the Introduction). In calculation of spectrum of excitations and of profiles of corresponding wave functions both types of deformations are practically equivalent. We use the Reissner-Nordstrom type deformation since it permits to receive in a natural way the scale of the "second" mass hierarchy, $\simeq 10^{-3}eV$.

To model the UV cut-off, i.e. compactification of the extra space, we introduce co-dimension one positive tension isotropic UV brane limiting the throat "from above". In \cite{Altsh7}, \cite{Altsh6} it was shown that for solution (\ref{2}) position of the isotropic positive tension UV-brane is stabilized by anysotropy of the Israel jump conditions at the top of the throat at $r = L/2^{1/3}$ where corresponding radion potential possesses stable extremum. Rolling down the radion potential, after stages of inflation and reheating Universe resides in this extremum in modern epoch. For the goals of the present paper we consider this stable state of the system and look only at the region inside the throat $r\ll L$ where expressions for metric and dilaton in (\ref{2}) take the form:

\begin{eqnarray}
\label{3}
&&ds_{(10)}^{2}=\left(\frac{r}{L}\right)^{9/8} \left [\tilde{g}_{\mu\nu}dx^{\mu}dx^{\nu}+  \left(\frac{T}{2\pi}\right)^{2} U d\theta^{2} \right]
\\  \nonumber
&& + \left(\frac{r}{L}\right)^{-15/8} \left[\frac{dr^{2}}{U}+r^{2}g_{ij}dy^{i}dy^{j}\right],
\qquad e^{\varphi}=g_{s}\left(\frac{r}{L}\right)^{3/4}.
\end{eqnarray}
$U(r)$ and $F_{(4)}$ are given in (\ref{2}), and from (\ref{2}), (\ref{3}) it is received $F_{(2)\theta r}=3\sqrt{2}\,M\,g_{s}^{-3/4}(T/2\pi)(l^{3}/r^{4})$. Isotropic coordinate $r$ varies between value $l$ determining the IR end (tip) and value $L$ approximately determining the UV end (top) of the throat: 

\begin{equation}
\label{4}
r_{IR}=l < r < L=r_{UV}.
\end{equation}
At the point $r=l$ where $U=0$ the co-dimension two brane may be located which results in appearance of conical singularity. We shall not consider this option and take the smooth end of the IR throat with zero deficit angle when period $T$ of cyclic coordinate is unambiguously determined from (\ref{3}):

\begin{equation}
\label{5}
T= \frac{2\pi}{3}\, L\, \left(\frac{L}{l}\right)^{1/2}.
\end{equation}

Physical values calculated in the paper will depend on two dimensionless parameters of the model: $ML$ and $L/l$. The value of $ML$ is limited from below since maximal (at $r=l$) space-time curvature $R^{(10)}$ must remain below the string scale $l_{s}^{-2}$. This gives:

\begin{equation}
\label{6}
ML > g_{s}^{-1/4}\left(\frac{L}{l}\right)^{1/16}.
\end{equation}

Proper distance $z$ along the throat (\ref{3}) quite slowly depends on the isotropic coordinate $r$:

\begin{equation}
\label{7}
z= \int_{l}^{r}N\,dr = \int_{l}^{r}\left(\frac{r}{L}\right)^{-15/16}\frac{dr}{\sqrt{U}} \cong 16\,L\, \left(\frac{r}{L}\right)^{1/16},
\end{equation}
cf. logarithmic dependence on $r$ in RS model or in the throat-like models of The Type IIB supergravity.

To receive from (\ref{1}) the D4 Einstein gravity Action it is necessary to integrate out six extra coordinates in the limits (\ref{4}) and with use of the background solution (\ref{3}). This gives D4 gravity Action $M^{8}V_{w}\int {\tilde R}^{(4)}\sqrt{-\tilde g^{(4)}}d^{4}x$ (${\tilde R}^{(4)}$ is curvature in D4 space-time with auxiliary metric ${\tilde g}_{\mu\nu}$, $V_{w}$ is so called "warped volume", symbolically $V_{w}=\int (\sqrt{-g^{(10)}}/b^{2}) d^{6}x$ \cite{Wolfe02}). After that the standard Einstein Action $M_{Pl}^{2}\int R^{(4)}\sqrt{-g^{(4)}}d^{4}x$ is received with a following scale transformation from the auxiliary 4-metric $\tilde{g}_{\mu\nu}=\eta_{\mu\nu}+\tilde{h}_{\mu\nu}(x)$ in (\ref{2}) or (\ref{3}) to the physical metric $g_{\mu\nu}$ (this transformation corresponds to the standard normalization of the fluctuations of metric zero mode): 

\begin{equation}
\label{8}
g_{\mu\nu}=\tilde{g}_{\mu\nu}\,\frac{M^{8}V_{w}}{M_{Pl}^{2}} = \tilde{g}_{\mu\nu}\, Z^{2},  \qquad  Z=\frac{M}{M_{Pl}}\, 7.4 \cdot (ML)^{3}\left(\frac{L}{l}\right)^{1/4},
\end{equation}
in the last equality expression (\ref{5}) for period of cyclic coordinate was used and we took compact subspace $K$ to be a four-sphere of unit radius.

Thus to get physical masses or lengths in the Einstein frame in four dimensions we must rescale by the factor $Z$ (\ref{8}) corresponding values in the auxiliary 4-dimensional space-time described by metric $\tilde{g}_{\mu\nu}$ in (\ref{2}), (\ref{3}). This gives for example the following expressions in Planck units for the physical proper length of the throat (\ref{7}) and for the physical value of period $T$ (\ref{5}):

\begin{equation}
\label{9}
z_{\it{phys}}=\int_{l}^{L}N\,dr \cdot Z=M_{Pl}^{-1} \cdot 10^{2} \cdot (ML)^{4} \left(\frac{L}{l}\right)^{1/4} \simeq M_{Pl}^{-1}\cdot 10^{10},
\end{equation}

\begin{equation}
\label{10}
T_{\it{phys}}=T \cdot Z =M_{Pl}^{-1} \cdot 15 \cdot (ML)^{4} \left(\frac{L}{l}\right)^{3/4} \simeq M_{Pl}^{-1}\cdot 10^{16}.
\end{equation}
The last numbers in (\ref{9}), (\ref{10}) are given for illustrative purpose for the values of dimensionless parameters $ML=10$ and $L/l=10^{15}$ which are at the brink of fulfillment of the low-energy condition (\ref{6}) and which are ajusted to make $T_{\it{phys}}^{-1}$ in (\ref{10}) of order $1$ TeV, i.e. of order of the electroweak scale (the reason is that period $T$, being the size of the largest extra dimension, determines the mass scale of the lower excitations, i.e. the mass gap, cf. \cite{Wolfe02}, and we want this scale to be the electroweak one).

From (\ref{3}) it is seen that size of cyclic dimension at the UV end of the throat given by the value of period (\ref{10}) is essentially above the proper length of the throat along the isotropic coordinate $r$ (\ref{9}). Thus the throat is really "fat". 

\section{Spectrum of scalar field. 
Natural appearance of the mass hierarchy}

We begin from the simplest and well known case of the bulk zero mass scalar field $\Phi (x^{A})$, $A=0,1...9$, which equation in space-time (\ref{2}) looks as:

\begin{equation}
\label{11}
\left[\frac{1}{b^{2}}\tilde{g}^{\mu\nu}\frac{\partial^{2}}{\partial x^{\mu}\partial x^{\nu} } + \frac{1}{c^{2}}\frac{\partial^{2}}{\partial \theta^{2}}+ \frac{1}{J}\frac{\partial}{\partial r}\left(\frac{J}{N^{2}}\frac{\partial}{\partial r}\right) + \frac{1}{a^{2}}\triangle^{(K)}\right] \Phi = 0,
\end{equation}
where $J=b^{4}cNa^{4}$, $\triangle^{(K)}$ is Laplacian in compact sub-space $K$. Standard separation of variables $\Phi =e^{ipx}\, e^{ik\theta}\, \phi_{q}(y^{j})\, f_{n,k,q}(r)$ ($p^{2}= - \widetilde{m}_{n}^{2}$, $n$ is spectral number, $n=1,2...$; $k=0,\pm 1,\pm 2...$; $\triangle^{(K)} \phi_{q}(y^{j}) = - \lambda_{q}\phi_{q}(y^{j})$) with use of metric factors $b, c, N, a$ from (\ref{2}), (\ref{3}) and expression (\ref{5}) for period $T$ gives following equation for the wave function $f(y)$ ($y\equiv r/l$, we omit here the eigenfunction indices):

\begin{equation}
\label{12}
(y^{7}-y) \frac{d^{2}f(y)}{dy^{2}}+ (4y^{6}+2)\frac{df(y)}{dy}
+\left(\mu^{2}y^{4}- \frac{9k^{2}y^{10}}{y^{6}-1}-\lambda y^{5}\right) f(y) = 0.
\end{equation}
Here dimensionless mass parameter $\mu$ was introduced:

\begin{equation}
\label{13}
\mu=\widetilde{m}L\left(\frac{L}{l}\right)^{1/2}.
\end{equation}
Thus with account of (\ref{8}), (\ref{13}) physical mass of the excitations in Einstein frame in four dimensions is in a following way expressed through $\mu$:

\begin{equation}
\label{14}
m_{\it{phys}}=\widetilde{m}\, Z^{-1}=M_{Pl}\,\frac{\mu}{7.4\, (ML)^{4}}\left(\frac{l}{L}\right)^{3/4}=2\mu \,T_{\it{phys}}^{-1}\cong \mu \cdot 1TeV,
\end{equation}
where expression for $T_{\it{phys}}$ is given in (\ref{10}), number in the last equality is valid for the choice of dimensionless parameters of the model $ML=10$, $L/l = 10^{15}$ (see comment to (\ref{10})). 

Spectrum of $\mu$ is found from the demand of regular behaviour of solution of Eq. (\ref{12}) at $y=1$ (IR end of the throat) and from the $Z_{2}$-symmetry condition $df/dy=0$ at the UV brane, i.e. at very big value of $y=L/l$. Direct calculations with use of Maple 9.5 show that there is a mass gap at $k=0$, $\lambda =0$ between the zero mode $\mu=0$, $f_{0}=const$ and first non-zero mode. We present the lowest non-zero values of three towers of $\mu_{n,k}^{2}$ for three "angular" modes $k=0, \, 1, \, 2$ (case $\lambda = 0$ in (\ref{12}) was considered, non-zero $\lambda$ will only increase the eigenvalues):

$$
k=0: \qquad \mu_{1,0}^{2}=6.3; \qquad  \mu_{2,0}^{2}=16.4; \qquad \mu_{3,0}^{2}=30.5
$$

$$
k=1: \qquad \mu_{1,1}^{2}=18.3; \qquad  \mu_{2,1}^{2}=30.5; \qquad \mu_{3,1}^{2}=46.3
$$

$$
k=2: \qquad \mu_{1,2}^{2}=47.1; \qquad  \mu_{2,2}^{2}=60.0; \qquad \mu_{3,2}^{2}=78.2
$$

Corresponding wave functions $f_{n,k}(y)$ oscillate in the region $1<y<10$, and all of them decrease rapidly, $f(y) \sim y^{-3}$, at larger $y$. Thus it is seen that profiles of massive excitations are essentially concentrated in vicinity of the IR end of the warped throat. As it was said in the Introduction gravitational accretion of massive matter to the IR end of the strongly warped space-time substitutes the trapping at the IR brane. "Life in a tip" is perhaps physically equivalent to "life on a brane", but is more natural from the theoretical point of view because does not suppose asymmetric treatment of bulk and brane matter, cf. \cite{Benson}.

These results are in no way new. Equations of type (\ref{12}), although slightly different because Schwarzschild type potential $U=1-l^{3}/r^{3}$ was used in (\ref{2}), were not a once studied earlier, see \cite{Witten}-\cite{Myers8} and references therein. Also it is well known that mass scale of the lowest KK excitations in these models is determined by $T^{-1}$, cf. (\ref{14}).

Dependence (\ref{14}) of physical mass of excitations on dimensionless parameters $ML$, $L/l$ is different in different theories and models of the throat. Let us look at the case of the $AdS_{5}\times S^{5}$ throat of the Type IIB supergravity. Inside the throat space-time metric is:

\begin{equation}
\label{15}
ds_{10}^{2}= \frac{r^{2}}{L^{2}}\tilde{g}_{\mu\nu}dx^{\mu}dx^{\nu} + \frac{L^{2}}{r^{2}}\left(dr^{2}+r^{2}d\Omega_{(5)}^{2}\right).
\end{equation}

We terminate this throat "by hand" at the UV and IR ends, $l<r<L$ like in (\ref{4}), where appropriate boundary conditions must be imposed upon the wave functions of KK excitations. 
For metric (\ref{15}) the analogous to (\ref{12}) equation for the zero-mass bulk scalar field is:

\begin{equation}
\label{16}
\left[y^{2}\frac{d^{2}}{dy^{2}}+5y\frac{d}{dy}+ \mu^{2}\right]\,f = 0,
\end{equation}
where again $y=r/l$ and in this model $\mu = \widetilde{m}L^{2}/l$ (cf. expression (\ref{13}) for parameter $\mu$ in the model of the Type IIA supergravity considered above). Proper boundary conditions at $y=1$ and at $y=y_{UV}=L/l \gg 1$ again give the lowest eigenvalues of $\mu$ of order $1 \div 10$. Calculating the "warped volume" from the higher-dimensional Action $M^{8}\int R^{(10)}...$ and repeating the logic of formulae (\ref{8}), (\ref{14})  gives in this case for the physical mass of excitations:

\begin{equation}
\label{17}
m_{n\it{phys}}=M_{Pl}\frac{\mu_{n}}{\sqrt{\Omega_{5}}}\,\frac{1}{(ML)^{4}}\,\frac{l}{L},
\end{equation}
where $\Omega_{5}$ is volume of 5-sphere of unit radius. Low energy string approximation makes sense at $ML > 10$. For $ML=10$ needed mass hierarchy $10^{-15}$ follows from (\ref{17}) at $L/l \approx 10^{10}$.

As it was noted in \cite{Wolfe02}, \cite{Chen}, the method of cutting IR and UV ends of the throat "by hand" is a simple way to calculate the spectrum of KK excitations at the background of Klebanov-Strassler throat \cite{Klebanov}. This essentially simplified consideration gives practically the same results as the exact one, performed for example in \cite{Firou}. Correspondence between lengths $l$, $L$ in (\ref{15})-(\ref{17}) and deformed conifold parameters is: $l=\epsilon ^{2/3}$, $L^{4}=4\pi g_{s}N_{c}l_{s}^{4}$; for KS throat $\Omega_{5}$ must be changed to the volume of base $T^{1,1}$ in the expression (\ref{17}) for eigenvalues of the mass spectra.

\section{Spectrum of Kaluza-Klein gauge fields.
\\ 
Higgs mechanism without Higgs scalar}

Conventional Kaluza-Klein approach considers gauge fields in 4 dimensions as a non-diagonal components of the higher-dimensional metric associated with isometries in extra space. In our case these are Young-Mills potential $B^{j}_{\mu}=B^{\gamma}_{\mu}\,\xi_{\gamma}^{j}(y)$ of the isometry group of compact subspace $K$ ($\xi_{\gamma}^{j}$ are normalized Killing vectors in $K$, $\gamma$ is the isometry group index) and abelian potential $B_{\mu}^{(\theta)}$ associated with cyclic coordinate $\theta$ in (\ref{2}). We denote corresponding non-abelian and abelian field strength as $F_{\mu\nu}^{\gamma}$ and $F_{\mu\nu}^{(\theta)}$ and consider these KK gauge fields dependent only on 4-d coordinates $x^{\mu}$ and isotropic coordinate $r$. Then reduction of Action (\ref{1}) to 5 dimensions with use of background (\ref{2}) gives following Actions $S^{(5)}_{(K)}[B^{\gamma}_{\mu}]$, $S^{(5)}_{(\theta)}[B^{(\theta)}_{\mu}]$ for the KK gauge fields $B^{\gamma}_{\mu}(x^{\nu},r)$ and $B_{\mu}^{(\theta)}(x^{\nu},r)$ (we take Minkowski 4-d space-time):
                                                                                                                                                                                                                                                                          
\begin{eqnarray}
\label{18}
&& S^{(5)}_{(K)}[B^{\gamma}_{\mu}]= M^{8}2\pi \Omega
\int d^{4}x \int_{l}^{L} \left[-\frac{1}{4}\frac{a^{2}}{b^{4}}F^{\gamma}_{\mu\nu}F^{\gamma \mu\nu}-  \right.  \nonumber
\\ 
&& \left.  -\frac{1}{2}\frac{a^{2}}{b^{2}N^{2}}B^{\gamma}_{\mu,r}B^{\gamma \mu}_{,r} - \frac{1}{2}\frac{a^{2}}{b^{2}}\frac{e^{\varphi/2}\bar F_{(4)}^{2}}
{4!}B^{\gamma}_{\mu}B^{\gamma \mu}\right]J\,dr =
\\
&& =M^{8}TL^{7}\Omega \int d^{4}x \int_{l}^{L} \left[-\frac{(F^{\gamma}_{\mu\nu})^{2}}{4} -\frac{(B^{\gamma}_{\mu,r})^{2}}{2}\left(\frac{r}{L}\right)^{3}U-\frac{(B^{\gamma}_{\mu})^{2}}{2}\frac{9r}{L^{3}}\right]\frac{dr}{L}, \nonumber
\end{eqnarray}

\
\

\begin{eqnarray}
\label{19}
&& S^{(5)}_{(\theta)}[B^{(\theta)}_{\mu}]=M^{8}2\pi \Omega \int d^{4}x \int_{l}^{L} \left[-\frac{1}{4}\frac{c^{2}}{b^{4}}F^{(\theta)}_{\mu\nu}F^{(\theta)\mu\nu}-\frac{1}{2}\frac{c^{2}}{b^{2}N^{2}}B^{(\theta)}_{\mu,r}B^{(\theta)\mu}_{,r}- \right.  \nonumber
\\ 
&& \left.  -\frac{1}{2}\frac{c^{2}}{b^{2}}\frac{e^{3\varphi/2}\bar F_{(2)}^{2}}{2M^{2}}
B^{(\theta)}_{\mu}B^{(\theta)\mu}\right] J\,dr=M^{8}\frac{T^{3}}{4\pi^{2}}L^{5}\Omega \int d^{4}x \int_{l}^{L} \left[-\frac{(F^{(\theta)}_{\mu\nu})^{2}}{4}\frac{r}{L}- \right.  \nonumber
\\ 
&& \left. -\frac{(B^{(\theta)}_{\mu,r})^{2}}{2}\left(\frac{r}{L}\right)^{4}U-\frac{(B^{(\theta)}_{\mu})^{2}}{2}\frac{18\,l^{6}}{L^{8}}\left(\frac{L}{r}\right)^{4}\right] U\frac{dr}{L}, 
\end{eqnarray}
where $a$, $b$, $c$, $N$, $\varphi$, $U$, $T$ are given in (\ref{2}), (\ref{3}), (\ref{5}), and $\bar F_{(4)}^{2}$, $\bar F_{(2)}^{2}$ are calculated with use of background solution (\ref{2}) i.e. in the absence of KK gauge fields $B_{\mu}^{\gamma}$, $B_{\mu}^{(\theta)}$; $J=b^{4}cNa^{4}$, $\Omega$ is volume of compact subspace $K$, comma means derivative over $r$. Last expressions in the RHS of (\ref{18}), (\ref{19}) are written inside the throat with use of (\ref{3}). Summing over group index $\gamma$ and contraction of $\mu, \, \nu$ with Minkowski metric $\eta_{\mu\nu}$ in the terms like $(F^{\gamma}_{\mu\nu})^{2}$ etc. are supposed.

Mass terms $(B^{\gamma}_{\mu})^{2}$ and $(B^{(\theta)}_{\mu})^{2}$ in (\ref{18}), (\ref{19}) result from the 4-form and 2-form terms in (\ref{1}) in direct analogy with Higgs mechanism - because of existence of non-zero matter fields $F_{(4)\,ijkl}$ and $F_{(2)\theta r}$ in background solution (\ref{2}). These fields substitute the charged scalar field condensate in the conventional Higgs mechanism. Appearance of the mass terms of gauge fields is seen from direct calculation of corresponding terms in (\ref{1}) when there are non-zero non-diagonal components $B^{\gamma}_{\mu}\xi^{i}_{\gamma}$, $B^{(\theta)}_{\mu}$ of the 10-dimensional metric:

\begin{equation}
\label{20}
F_{(4)ABCD}F_{(4)}^{ABCD}= \bar F_{(4)}^{2}\left[1+\frac{a^{2}}{b^{2}}\tilde{g}^{\mu\nu}B^{\gamma}_{\mu}B^{\delta}_{\nu}\xi_{\gamma}^{i}\xi_{\delta}^{j}g_{ij}+ B^{4}+B^{6}+B^{8}\,\right],
\end{equation}

\begin{equation}
\label{21}
F_{(2)AB}F_{(2)}^{AB}= \bar F_{(2)}^{2}\left[1+\frac{c^{2}}{b^{2}}\tilde{g}^{\mu\nu}B^{(\theta)}_{\mu}B^{(\theta)}_{\nu}\right].
\end{equation}
Since linear equations are considered the higher order in $B_{\mu}^{\gamma}$ terms shown symbolically in (\ref{20}) are non essential here. 

The question may arise why quantities (\ref{20}), (\ref{21}) which look like D10 scalars are not gauge invariants (in KK picture gauge transformations are just a special coordinate transformations in higher dimensions)? Like in standard Higgs mechanism the reason is in non-invariance of the ground state, in our case - non-invariance of the background solution (\ref{2}). In calculating (\ref{20}), (\ref{21}) we used D10 metric $g^{AB}$ modified by the non-diagonal terms $B_{\mu}^{i}$, $B_{\mu}^{\theta}$ whereas we took the non-modified background values of $F_{(4)ijkl}$, $F_{(2)\theta r}$ given in (\ref{2}). Situation literarty repeats the appearance of the gauge-non-invariant gauge field mass term from the gauge-invariant Lagrangian of the charged Higgs scalar field. Under gauge (i.e. corresponding higher-dimensional coordinate) transformations 4-form and 2-form fields of background (\ref{2}) acquire components "living" in "our" four-dimensional space-time; the choice of the non-modified ground state values $F_{(4)ijkl}$ and $F_{(2)\theta r}$ in calculating quantities (\ref{20}), (\ref{21}) is in a sense in parallel with the choice of the unitary gauge in Higgs theory.

Our goal is to find spectra of gauge potentials $B^{\gamma}_{\mu}$, $B^{(\theta)}_{\mu}$ which dynamics is governed by the Actions in RHS of (\ref{18}), (\ref{19}). Separation of variables: 

\begin{equation}
\label{22}
B^{\gamma}_{\mu}(x^{\nu},r)=\Sigma_{n}B^{\gamma}_{\mu n}(x^{\nu})f^{(K)}_{n}(r),  \qquad B^{(\theta)}_{\mu}(x^{\nu},r)=\Sigma_{n}B^{(\theta)}_{\mu n}(x^{\nu})f^{(\theta)}_{n}(r),
\end{equation}
where $B^{\gamma}_{\mu n}(x^{\nu})$ and $B^{(\theta)}_{\mu n}(x^{\nu})$ are gauge fields in four dimensions of masses $\widetilde{m}_{(K)n}$ and $\widetilde{m}_{(\theta) n}$ correspondingly, gives finally the following equations for the wave functions $f^{(K)}(y)$ and $f^{(\theta)}(y)$ ($y=r/l$, spectral index $n$ is omitted):

\begin{equation}
\label{23}
(y^{7}-y) \frac{d^{2}f^{(K)}(y)}{dy^{2}}+ (3y^{6}+3)\frac{df^{(K)}(y)}{dy}+\left(\mu_{(K)}^{2}y^{4}- 9y^{5}\right) f^{(K)}(y) = 0.
\end{equation}

\begin{equation}
\label{24}
(y^{8}-y^{2}) \frac{d^{2}f^{(\theta)}(y)}{dy^{2}}+ (4y^{7}+8y)\frac{df^{(\theta)}(y)}{dy}+\left(\mu_{(\theta)}^{2}y^{5}- 18 \right) f^{(\theta)}(y) = 0.
\end{equation}

These equations resemble similar equation (\ref{12}) (in case $k=0$, $\lambda =0 $ in (\ref{12})) for the bulk scalar field. However in (\ref{12}) there are no bulk mass term of scalar field. Contrary to it "Higgs" terms $9y^{5}f^{(K)}$ in (\ref{23}) and $18f^{(\theta)}$ in (\ref{24}) appeared from the mass terms in Actions (\ref{18}), (\ref{19}). Dimensionless parameters $\mu_{(K)}$, $\mu_{(\theta)}$ in (\ref{23}), (\ref{24}) are connected with corresponding masses $\widetilde{m}_{(K)}$, $\widetilde{m}_{(\theta)}$ in the same way like in (\ref{13}).

At $y\gg 1$ solutions of every of Eq-s (\ref{23}), (\ref{24}) possess two branches: growing and decreasing for $f^{(K)}$ and constant and decreasing for $f^{(\theta)}$:

\begin{equation}
\label{25}
f^{(K)}(y)= C_{1}y^{\sqrt{10}-1}+C_{2}y^{-\sqrt{10}-1}, \qquad f^{(\theta)}(y)= C_{3}+\frac{C_{4}}{y^{3}}.
\end{equation}

Like in scalar field case of Sec. 3 mass spectrum is determined from the demand of regular behavior of solution at $y=1$ combined with the $Z_{2}$-symmetric boundary condition at the UV end: $df/dy=0$ at $y=L/l \gg 1$. But now there are no zero modes which satisfy these demands.
Number calculations of spectrum of normalized massive modes of both equations (\ref{23}), (\ref{24}) give similar to the scalar case of Section 3 results for eigenvalues:

three lowest eigenvalues of Eq. (\ref{23}) are:

$$
\mu_{(K)}^{2}=17.2,\,\,34.8,\,\,55.6,
$$

and three eigenvalues of Eq. (\ref{24}) are: 

$$
\mu_{(\theta)}^{2}=9.6,\,\, 21.7, \,\, 37.7.
$$

Physical masses of excitations of KK gauge fields $B^{\gamma}_{\mu}$, $B^{(\theta)}_{\mu}$ in Einstein frame in four dimensions are determined through corresponding parameters $\mu$ by expression (\ref{14}) and are of order of electroweak scale for the proper choice of $ML$ and $L/l$ (e.g. for $ML = 10$, $L/l = 10^{15}$). Wave function $f_{n}(y)$ of the $n$-th massive mode possesses $n$ oscillations and is concentrated in the vicinity of the IR end of the throat. There is however the special eigenmode of Eq. (\ref{24}) with extremely small eigenvalue $\mu_{(\theta)0}$ which is diluted in extra space, see next Section.

Young-Mills KK gauge field associated with isometry group of compact subspace $K$ acquires mass because we took magnetic fluxbrane solution (\ref{2}) where 4-form "lives" at $K$ ($F_{ijkl} \ne 0$). If the electric fluxbrane solution was considered then Higgs-type mechanism (\ref{20}) will not work, term $9y^{5}f^{(K)}$ will be absent in Eq. (\ref{23}) and non-abelian KK gauge field $B^{\gamma}_{\mu}$ will possess massless mode in 4 dimensions which wave function is constant in extra space.

To conclude this Section we note that abelian gauge field $F_{(2)AB}$ of the Type IIA supergravity Action (\ref{1}) may be also considered as a KK gauge field received by compactification of periodic 11-th dimension of D11 M-theory; dilaton $\varphi$ in this case charachterizes the size of the compactified 11-th dimension \cite{SUGRA}. We did not put down equation for excitations of this gauge field since for the background (\ref{2}) this equation is identical with equation (\ref{12}) for excitations of bulk zero mass scalar field. The reason for this coincidence is in the special feature of solution (\ref{2}) which gives similar dependence on $r$ of factors $e^{3 \varphi /2}$ and $b^{2}$ determining corresponding equations. Thus spectrum of excitations of this gauge field is given in Section 3, including of course the zero mode which is constant in extra space.

\section{Appearance of "second mass hierarchy"}

There are simple exact solutions for zero modes of Eq. (\ref{24}), i.e. in case $\mu_{(\theta)}=0$ in (\ref{24}). One of them ($f=y^{3}/(y^{6}-1)$) being singular at the IR end ($y=1$) is not physical. The other one, regular at $y=1$, looks as:

\begin{equation}
\label{26}
f^{(\theta)}_{0}(y)= \frac{y^{3}}{y^{3}+1}.
\end{equation}
This solution however does not meet the demand of $Z_{2}$-symmetry ($df/dy=0$) at the UV end of the throat, i.e. at $y=L/l$. However deviation of $df^{(\theta)}_{0}/dy$ from zero at this point is quite small ($= 3y^{-4}=3(l/L)^{4}\ll 1$). Because of it $Z_{2}$-symmetry may be restored with introduction of small mass parameter $\mu_{(\theta)0}$ in (\ref{24}). It is easy to receive from (\ref{24}) solution $f^{(\theta)}_{0\it{corr}}$ corrected with introduction of small mass. Thus instead of (\ref{26}) we have in the first order in $\mu_{(\theta)0}^{2}$:

\begin{equation}
\label{27}
f^{(\theta)}_{0\it{corr}}(y)= \frac{y^{3}}{y^{3}+1}\left[1+\frac{\mu_{(\theta)0}^{2}}{2y}\right].
\end{equation}

It is seen that $Z_{2}$-symmetry condition $df^{(\theta)}_{0\it{corr}}/dy=0$ at $y=L/l$ is fulfilled for $\mu_{(\theta)0}=\sqrt{6}l/L$ which is $\ll 1$ (for "ordinary" massive modes of Sections 3, 4 we had parameter $\mu$ of order ten). Substitution of this value of $\mu_{(\theta)0}$ into Eq. (\ref{14}) gives the following value of physical mass $m_{(\theta)0}$:

\begin{equation}
\label{28}
m_{(\theta)0}= M_{Pl}\frac{1}{(ML)^{4}}\left(\frac{l}{L}\right)^{7/4}\cong 10^{-3}eV.
\end{equation}
Again last figure is received for illustrative purposes for values of parameters $ML=10$, $L/l = 10^{15}$.

The logic of appearance here of the small mass scale ("second mass hierarchy") is simple. Magnetic monopole abelian field $F_{(2)\theta r}$ in (\ref{2}) responsible for the Reissner-Nordstrom term $l^{6}/r^{6}$ in potential $U(r)$ in (\ref{2}), (\ref{3}) quickly decreases with increase of $r$: $F_{(2)\theta r} \sim (L/r)^{4}$, as it is seen from (\ref{2}), (\ref{3}). Because of it mass term $(B^{(\theta)}_{\mu})^{2}$ in  (\ref{19}) is concentrated in vicinity of the IR end of the throat. Its "tail" at the UV end is quite small, hence violation of the $Z_{2}$-symmetry of zero mode (\ref{26}) produced by this "tail"  is small as well. Thus mass of the gauge field which corrects this violation is also very small, like in (\ref{28}).

\section{Fabric of the gauge coupling constants}

In the reduced to 4 dimensions dynamical equations of the charged (i.e. 
depending on compact extra 
coordinates) modes of bulk matter fields 
$\psi$ the non-diagonal components of the higher-dimensional 
metric appear through "long derivatives" $\partial \psi/ \partial x^{\mu} - 
B_{\mu}\psi$ (symbolically, omitting group indices). We take normalization of 
eigenfunctions of gauge potentials (\ref{22}) in a way (also symbolically) 

\begin{equation}
\label{29}
\frac{\int \overline{\psi}f_{n}(r)\psi \,dr}{\int\overline{\psi}\psi \,dr}=1,
\end{equation}
which preserves in 4 dimensions the form ($\partial_{\mu}- B_{\mu})$ of the long 
derivative when value of the 4-d gauge coupling constant $g$ is given by the 
coefficient $1/4g^{2}$ before the $F^{2}_{\mu\nu}$ term in the gauge field Lagrangian. 

Condition (\ref{29}) means that for the constant in extra space zero modes of 
gauge fields ($f=const$ guarantees the universality of the long-range gauge 
interactions, i.e. its independence on profiles of matter fields $\psi(r)$) 
wave functions are normalized to 1: $f^{(K)}_{0}=1$, $f^{(\theta)}_{0}=1$. 
And this condition also means that for massive modes of Eqs. (\ref{23}), 
(\ref{24}) we take normalization when $f^{(K)}_{n}(y), f^{(\theta)}_{n}(y) \cong 1$ in vicinity of the IR end of the throat ($y=r/l \cong 1$) where also the wave functions $\psi (r)$ of the charged massive matter fields are concentrated.

Let us calculate values of gauge coupling constants $g_{(K)n}$, $g_{(\theta)n}$, $g_{(11)n}$ in 4 dimensions of different modes of the gauge fields $B^{\gamma}_{\mu}$, $B^{(\theta)}_{\mu}$ (\ref{22}) and of the abelian gauge potential $A_{\mu}(x^{\nu}, r)=\Sigma_{n}A_{\mu n}(x^{\nu})f^{(11)}_{n}(r)$ in (\ref{1}). Gauge potential $A_{\mu}$, as it was noted in the end of Sec. 4, may be considered as non-diagonal component of D11 metric of M-theory compactified to 10 dimensions, $f^{(11)}_{n}(r)$ are profiles in extra space of its eigenmodes. When normalization conditions (\ref{29}) are imposed expressions for these coupling constants are received by intergration over $r$ of the $(F_{\mu\nu})^{2}$ terms in the RHS of (\ref{18}), (\ref{19}) and by integration over 6 extra coordinates of the $(F_{(2)\mu\nu})^{2}$ term in (\ref{1}):

\begin{equation}
\label{30}
\frac{1}{4g_{(K)n}^{2}}= 55\cdot (ML)^{8}\left(\frac{L}{l}\right)^{1/2}\int_{l}^{L}f^{(K)\,2}_{n}(r)\,\frac{dr}{L},
\end{equation}

\begin{equation}
\label{31}
\frac{1}{4g_{(\theta)n}^{2}}= 6\cdot (ML)^{8}\left(\frac{L}{l}\right)^{3/2}\int_{l}^{L}f^{(\theta)\,2}_{n}(r)\, U(r)\frac{r}{L}\, \frac{dr}{L},
\end{equation}

\begin{equation}
\label{32}
\frac{1}{4g_{(11)n}^{2}}= 55\cdot (ML)^{6}\left(\frac{L}{l}\right)^{1/2}g_{s}^{3/2}\int_{l}^{L}f^{(11)\, 2}_{n}(r)\, \frac{r}{L}\,\frac{dr}{L},
\end{equation}
where formula (\ref{5}) for period $T$ was used and we took volume $\Omega$ of the compact subspace $K$ equal to volume of 4-sphere of unit radius.

It is seen that for zero modes when profiles $f_{0}=1$, see (\ref{29}), integrals in the RHS of (\ref{30})-(\ref{32}) are of order one. Thus, for example, for the values of parameters $ML=10$, $L/l=10^{15}$ used in (\ref{14}) expressions (\ref{30})-(\ref{32}) give in this case extremely small values of the KK gauge coupling constants: $g_{(K)0}\cong 10^{-9}$, $g_{(\theta)0}\cong 10^{-16}$, $g_{(11)0} \cong 10^{-8}\cdot g_{s}^{-3/4}$. The same follow from (\ref{31}) for the "corrected" zero mode (\ref{27}) which is equal to 1 practically everywhere in the throat.

On the other hand for higher modes of the gauge fields, when profiles $f_{n}(r)$ are concentrated at the IR end of the warped space-time and their normalization is taken like in (\ref{29}), integrals in the RHS of (\ref{30})-(\ref{32}) essentially suppress the result; for $n\ge 1$ these expressions give:

\begin{equation}
\label{33}
\frac{1}{4g_{(K)n}^{2}}\approx \frac{1}{4g_{(K)0}^{2}}\cdot\frac{l}{L}, \quad \frac{1}{4g_{(\theta)n}^{2}}\approx \frac{1}{4g_{(\theta)0}^{2}}\cdot \left(\frac{l}{L}\right)^{2}, \quad \frac{1}{4g_{(11)n}^{2}}\approx \frac{1}{4g_{(11)0}^{2}}\cdot \left(\frac{l}{L}\right)^{2}.
\end{equation}
Thus for the calculated above extremely small values of the zero modes' 
coupling constants $g_{0}$ the physically sensible values of the gauge coupling constants of massive modes ($n\ge 1$) are received from (\ref{33}): $g_{(K)n}=g_{(K)0}(L/l)^{1/2}\cong 10^{-1}$ and $g_{{\theta}n}=g_{(\theta)0}(L/l) \cong 10^{-1}$. At the same time interaction of the masssive modes of supergravity abelin field $A_{\mu}$ proves to be too strong: $g_{(11)n}=g_{(11)0}(L/l) \cong 10^{7}g_{s}^{-3/4}$.

The problem of KK gauge coupling constants outlined here is of general nature. As it was said the universality of the long-range gauge interaction is guaranteed only in case wave function of the gauge field zero mode is constant in extra space. Thus value of the gauge coupling constant is determined by the size of extra space.
In the supergravity-based models presence of additional extra dimensions, especially of the cyclic one with period of the electroweak scale (see (\ref{10})), catastrophically decreases long range gauge coupling constants and makes questionable the classical KK approach to the gauge fields in frames of compactified higher-dimensional supergravity models.

This problem exists even in absence of the additional cyclic dimension, in the throat-like models of the Type IIB supergravity, like model (\ref{15}) or model of KS throat \cite{Klebanov}. In the $AdS_{5} \times S^{5}$ model (\ref{15}) gauge coupling constant $g_{(5)0}$ of zero modes of KK gauge fields associated with isometry group of 5-sphere is determined from the expression:

\begin{equation}
\label{34}
\frac{1}{4g_{(5)0}^{2}}= (M_{(10)}L)^{8}\Omega_{5} \ln \frac{L}{l},
\end{equation}
where $M_{(10)}$ is Planck mass in $D10$ and $\Omega_{5}$ is volume of 5-sphere of unit radius. Low energy string approximation demands $M_{(10)}L \gg 1$ which makes KK gauge coupling $g_{(5)0}$ extremely small. 
The same (i.e. $g_{(5)n}\ll 1$ for $n\ge 1$) is true in this model for the gauge coupling constants of massive modes of gauge fields since $\int f_{n}^{2}dr/r$ which determines values of $1/4g_{(5)n}^{2}$ depends only "logarithmically" on the form of profiles $f_{n}(r)$.

At this point throat-like solutions of the Type IIB supergravity equations crucially differ from the corresponding backgrounds in the Type IIA supergravity where KK gauge coupling constants of higher modes given by expressions of type (\ref{33}) prove to be of the physically sensible values.

\section{Discussion}

\qquad In the final notes to \cite{SUGRA} it is said that in the 10-dimensional 
supergravity questions about observed physical phenomena may be translated 
into questions about properties of compact extra manifold. The toy model 
considered in the paper tries to follow this trend although it is far from 
ambitions to describe real world. We did not consider supersymmetry and its 
breakdown, group content of Standard Model, generations and appearance of 
their masses, etc. To find out to what extent the throat-like 
background of the supergravity theory compactified to four dimensions
may be used for description of the real world is a task for future work.
In particular it would be important to consider the possibility of 
generating of Fermions' masses with a "flux-Higgs" mechanism considered 
above.

The sub-eV mass scale received 
in the paper in the spectrum of KK gauge excitations with use of the 
"flux-Higgs" mechanism applied to the RN-deformed throat does not 
come in contradiction with experiment because of the extremely small value 
of corresponding gauge coupling constant (see Sec. 6). It would be 
more interesting to study the possibility 
to get in the same way the sub-eV mass of 
neutrino. It must be noted that in the Randall-Sundrum type models 
attempts to receive sub-eV scale in spectrum of 
gravitino excitations \cite{Pomarol}, \cite{ChenDark} or as the 
supersymmetry-breaking scale in the bulk \cite{Burgess04} were 
undertaken earlier.

As it was noted in the end of Sections 3, 6, the approach of the paper may be 
applied also to the throat-like models of the Type IIB supergravity. 
In particular the analogous to (\ref{19}) D5 theory of the KK gauge fields 
of the isometry group 
of $S^{5}$ for the model (\ref{15}) (or of the isometry group of base 
$T^{1,1}$ for KS throat) may be easily received with reduction from 10 
to 5 dimensions. 

Then questions arise: What may be the dual CFT theories at 
the 4-dimensional boundary if D5 gauge theories of type (\ref{18}), (\ref{19}) 
(or corresponding theories recived from backgrounds of the Type IIB 
supergravity) are considered as their supergravity duals in frames of the 
holographic approach? What are the quantum currents in dual D4 CFT which external 
sources are boundary values of D5 KK gauge potentials (\ref{22})?

In the holography business there are two complementary approaches to build 
the D5 gravity theory dual to D4 QCD \cite{Karch06}: the "bottom-up" 
ajustment of the D5 theory to the known properties of the low-dimensional 
CFT \cite{12}-\cite{14} or receiving of D5 theory with compactification 
from 10 dimensions, like e.g. it is done in \cite{HorPolch} where the 
starting theory is the gauge theory in D10 for an $SU(N_{c})$ gauge field 
and a 16 component Majorana-Weyl spinor. 
Thus it would be interesting to study if the D5 theories of 
type (\ref{18}), (\ref{19}) received from Type IIA 
or Type IIB supergravities as starting theories may be used in frames of holography approach 
for decription of 
connected Green's functions and properties of bound states of the realistic
CFT in 4 dimensions.

And there is another more general question about possible correspondence of 
classical KK-compactification approach and the dual holography 
approach to the low-dimensional theories. Matter content of D4 theory 
received in KK approach, i.e. 
with intergration out of extra dimensions, is given by the normalized in 
extra space modes of bulk fields. Whereas in the holography approach 
the boundary values 
of the non-normalized modes generate quantum currents of dual CFT; 
supergravity action considered as functional of these boundary values is 
used for calculation of spectra, decay constants etc. of bound states of the 
low-dimensional dual CFT. Perhaps the time came to remember Schwinger's 
"Source theory" with its idea of identification of particles (quantum fields) 
and their sources (quantum currents) \cite{Schwinger}. 

Natural realization of this idea (which is in line with "bootstrap" 
ambition to consider particles as bound states in $S$-matrix amplitudes 
of the same particles) could be possibly the {\it{bootstrap 
holography principle}} demanding the identity of two theories in 4 
dimensions: the KK one received by classical compactification of the normalized 
modes of bulk fields and the holography dual one describing the connected 
Green's functions and bound states {\bf {of the same theory}}. And may be 
this bootstrap holography principle will serve a sort of selection rule for 
the choice from plethora of possibilities of the higher-dimensional 
string-supergravity background vacuum state.

\section*{Acknowledgements} Author is grateful for fruitful criticism to 
participants of the Quantum Field Theory Seminar in the Theoretical Physics 
Department, Lebedev Physical 
Institute. This work was supported by the Program for Supporting Leading 
Scientific Schools (Grant LSS-1615.2008.2).

\end{document}